\documentclass[graybox]{svmult}

\usepackage{mathptmx}       
\usepackage{helvet}         
\usepackage{courier}        
\usepackage{type1cm}        
\usepackage{makeidx}         
\usepackage{graphicx}        
\usepackage{multicol}        
\usepackage[bottom]{footmisc}

\makeindex             

\begin{document}

\title{Gravitational Collapse and Disk Formation in Magnetized Cores}

\titlerunning{Gravitational Collapse and Disk Formation} 

\author{Susana Lizano and Daniele Galli}
\authorrunning{S. Lizano \& D. Galli} 

\institute{Susana Lizano \at Centro de Radioastronom\'\i a y
Astrof\'\i sica, UNAM, Apartado Postal 3-72, 58089 Morelia,
Michoac\'an, M\'exico \\
\email{s.lizano@crya.unam.mx}
\and Daniele Galli \at INAF-Osservatorio Astrofisico di Arcetri,
Largo E. Fermi 5, I-50125, Firenze, Italy \\
\email{galli@arcetri.astro.it}}

\maketitle

\abstract{We discuss the effects of the magnetic field observed
in molecular clouds on the process of star formation, concentrating
on the phase of gravitational collapse of low-mass dense cores,
cradles of sunlike stars.  We summarize recent analytic work and
numerical simulations showing that a substantial level of magnetic
field diffusion at high densities has to occur in order to form
rotationally supported disks. Furthermore, newly formed accretion
disks are threaded by the magnetic field dragged from the parent
core during the gravitational collapse.  These disks are expected to
rotate with a sub-Keplerian speed because they are partially supported by magnetic
tension against the gravity of the central star. We discuss how
sub-Keplerian rotation makes it difficult to eject disk winds and
accelerates the process of planet migration. Moreover, magnetic
fields modify the Toomre criterion for gravitational instability
via two opposing effects: magnetic tension and pressure increase
the disk local stability, but sub-Keplerian rotation makes the disk
more unstable. In general, magnetized disks are more stable than
their nonmagnetic counterparts; thus, they can be more massive and
less prone to the formation of giant planets by gravitational
instability.}

\section{Introduction}
\label{sec:Intro}

The goal of this review is to summarize recent theoretical work addressing
the role of magnetic fields in the process of star formation, and, in particular,
on the formation of circumstellar disks. In fact, the interstellar magnetic 
field dragged in the star plus disk system by the collapse of the parent cloud 
affects in the first place the process of disk formation itself (Section~\ref{sec:magbrak}), but also provides 
a natural mechanism for disk viscosity and resistivity via the MRI instability (Section~\ref{sec:Bdisk}),
affects the rotation curve of the disk (Section~\ref{sec:rotcurve}), its
stability properties (Section~\ref{sec:stabdisks}), and the migration of planets (Section~\ref{sec:migraplan}).
Unfortunately, the detection of magnetic fields in circumstellar disks is 
still an observational challenge (Section~\ref{sec:ossdisks}). For recent comprehensive 
reviews  see, e.g., K\"onigl \& Salmeron~(2011) for the role of magnetic fields in the 
process of disk formation, and Armitage~(2011) for the evolution of protoplanetary disks.

\section{Magnetic Fields in Molecular Clouds}
\label{sec:molclouds}

Theoretical considerations (Chandrasekhar \& Fermi 1953, Mestel \&
Spitzer 1956) show that a cloud of mass $M$ enclosing a magnetic
flux $\Phi$ can be supported by the magnetic field against its
self-gravity provided its non-dimensional mass-to-flux ratio $\lambda$
expressed in units of $(2\pi G^{1/2})^{-1}$ where $G$ is the
gravitational constant, is less than unity,
\begin{equation}
\lambda \equiv 2\pi G^{1/2} \left(\frac{M}{\Phi}\right) <1.
\end{equation}
Subcritical clouds with $\lambda < 1$ evolve on a timescale that
characterizes the diffusion of the magnetic field (e.g., Nakano
1979).  On the other hand, supercritical clouds with $\lambda  >
1$ cannot be supported by the magnetic field alone, even if the
field were perfectly frozen in the gas.  These clouds would collapse
on a magnetically diluted free-fall timescale.

A large
number of measurements of the intensity of the magnetic field has been obtained
for different ISM conditions with various techniques (see e.g. Crutcher~2012 
and the chapter by Crutcher, this volume).
The available measurements of magnetic field strength in molecular
clouds support the conclusion that on average, clouds are close to
the critical value $\lambda \approx 1$. OH Zeeman measurements at cloud
densities $\sim 10^3$ cm$^{-3}$ give mean mass-to-flux ratios
$\lambda \sim 2$--3 (Crutcher \& Troland 2008). In dense cores, the
mean value of the mass-to-flux ratio from CN Zeeman observations
is $\lambda  \sim 2$. Nevertheless, due to uncertainties in the
measurement of the gas column densities of a factor $\sim 2$, a
possible range of mass-to-flux ratios is $\lambda \sim 1$--4 (Falgarone
et al. 2008).  As we discuss below, these observed magnetic fields
are dynamically important for the cloud evolution and gravitational
collapse to form stars. In fact, these fields are well ordered on
the large scales of molecular clouds (e.g., Goldsmith et al. 2008).

Thanks to the presence
in the ISM of aspherical dust grains aligned with the field, polarization maps
of the thermal emission of molecular clouds at submillimeter wavelengths 
have made possible to determine the field geometry and the relative importance
of the ordered and turbulent components of the magnetic field.
At the core scales,  recent SMA maps of polarized dust continuum
emission show the hourglass morphology expected for a magnetically
controlled collapse rather than a disordered field dominated by
important levels of turbulent motions (e.g., Girart et al. 2006,
2009; Gon{\c c}alves et al.~2008, Tang et al. 2009). Moreover, the
statistical analysis of submillimeter polarization maps based on
the dispersion function method, shows that the ratio of the rms
turbulent component of the magnetic field to the mean value is of
the order of 0.1--0.5 (Hildebrand et al. 2009; Houde et al. 2009).
Thus, turbulent motions do not dominate the large scale well ordered
magnetic field.

As a first approximation, the magnetic field in molecular clouds
can be considered frozen to the gas. Even though the gas is is
lightly ionized, with an ionization fraction $\sim 10^{-8}$--$10^{-7}$
(see e.g. Caselli et al.~1998), collisions between charged particles
and neutrals efficiently transmit the Lorentz force to the largely
neutral gas. In these conditions, the relevant mechanism of field
diffusion is ambipolar diffusion (hereafter AD), originally proposed
by Mestel \& Spitzer~(1956), a process by which the fluid of charged
particles attached to the magnetic field can slowly drift with
respect to the fluid of neutral particles (see chapter by Zweibel in this 
volume). The field is then left
behind  with respect to the neutral gas and the mass-to-flux ratio
increases as the cloud condenses under the influence of its
self-gravity. Thus, an initially magnetically subcritical region
of mass sufficiently high for its self-gravity to overcome the support 
provided by thermal pressure and turbulent motions, evolves under
this process toward a centrally condensed core with supercritical 
mass-to-flux ratio, $\lambda \ge 1$ that eventually collapses and fragments
(e.g., Lizano \& Shu 1989; Tomisaka et al. 1990).  These authors
showed that, for typical conditions of molecular clouds (densities
$\sim 10^3$~cm$^{-3}$, magnetic fields $B \sim 30$~$\mu$G), the
AD timescale for core formation was a few $\times 10^6$~yr. Once
the dense cores are formed, with densities of $\sim 3 \times 
10^4$~cm$^{-3}$, they quickly evolve toward the stage of gravitational
collapse in a few $\times 10^5$~yr, of the order of the free-fall
timescale. The mass-to-flux ratio increases little in this condensation
process of core formation, thus, the major phase of flux loss occurs
at higher densities. At core densities, also Ohmic losses are far
too small to significantly reduce the magnetic field strength.  For
example, for an ionization fraction $\sim 10^{-8}$ and density $\sim
10^5$~cm$^{-3}$, the Ohmic dissipation time of a magnetic field
extending on a length scale of $\sim 0.1$~pc is of the order of
$10^{15}$~yr.  In contrast, at higher density, $n \sim 10^{11}$
cm$^{-3}$, characteristic of circumstellar disk formation, the field
can be effectively removed from the system by processes like Ohmic
diffusion and the Hall effect (see, e.g. Pinto et al.~2008; Pinto
\& Galli~2008). At some point in the star formation process magnetic
field removal is necessary to prevent the formation of stars with
MG fields, as would be the case under field freezing conditions.
This has been called the ``magnetic flux'' problem by Mestel \&
Spitzer (1956). Current observations show that at the surface of
young stars the fields have magnitudes  of kG (e.g., Johns-Krull
et al.~1999, 2004), they are likely to be generated by dynamo action,
rather than being fossil fields.

It has been under debate whether the formation and evolution of the
dense cores is controlled by magnetic fields as discussed above
(e.g., Mouschovias \& Ciolek~1999; Adams \& Shu~2007; Nakamura \& Li 2008) 
or by gravo-turbulent fragmentation driven by supersonic turbulence
(e.g., Padoan et al.~2001; Klessen et al.~2005; V\'azquez-Semadeni et al.~2005),
or by hierarchical gravitational collapse of a cloud as a whole 
(e.g., Ballesteros-Paredes et al. 2011a,b and chapter by
V\'azquez-Semadeni in this volume).  We
favor the first process because the large scale magnetic fields
dominate over their turbulent components and they are strong enough
to contribute to the support of a cloud core against its self-gravity 
if $\lambda$ is of order unity.  Furthermore,
as we will discuss below, these fields influence the dynamics of
the gravitational collapse and they are difficult to get rid of.

The process of gravitational collapse therefore separates logically
into two phases: ({\it a}\/) how cloud cores that were initially
sub-critical evolve to a state of being super-critical, and ({\it
b}\/) how cloud cores of both low and high mass that are super-critical
subsequently gravitationally collapse and possibly fragment once
they pass beyond the threshold of stability.  In this chapter we
will focus on the second problem, the phase of gravitational collapse
of a super-critical magnetized rotating core to form a star-disk
system.

\section{Gravitational Collapse of Magnetized Cores}
\subsection{Catastrophic Magnetic Braking and Disk Formation}
\label{sec:magbrak}

Recently, several studies have addressed the gravitational collapse
of a magnetized rotating cloud core to form a protostar plus disk
system. These studies have considered the ideal magnetohydrodynamic
(MHD) regime where the magnetic flux is frozen to the fluid, and
the non-ideal MHD regime where several diffusive processes are
taken into account. In particular, in their seminal work, Allen et al.
(2003) followed the ideal MHD collapse of a core threaded by a
large scale poloidal field, taking as an initial state a singular
isothermal  uniformly rotating toroid (Li \& Shu~1996). They found
a pseudo-disk predicted by Galli \& Shu~(1993a,b), a non-equilibrium
flattened structure formed around the star by the tendency of the
gas to flow along field lines and by the pinching Lorentz force
resulting from the bending of the field dragged to the central star.
These simulations also produced the slow outflows (with velocities
of few km~s$^{-1}$) found in previous numerical studies (e.g,
Tomisaka~2002). Nevertheless, the simulations did not show the
formation of a rotationally supported disk (RSD). Allen et al.
argued that RSDs do not form in the ideal MHD regime because the
enhanced strength and increased lever arm of the magnetic 
field dragged into the center of collapse results in a very efficient
transfer of angular momentum from the accretion region to the cloud envelope.
Subsequent numerical simulations confirmed that RSDs naturally form
if $B=0$ but do not form in strongly magnetized clouds (e.g., Price
\& Bate~2007, Fromang et al.~2006, Duffin \& Pudritz~2009).  In
fact, the process of magnetic braking was known to provide a way
to remove the cloud angular momentum and to allow the formation of
disks and binary stars (e.g., Mouschovias \& Paleologou~1980).
However, the braking found in these simulations of the gravitational
collapse phase was too efficient and prevented altogether the
formation of a disk.

\begin{figure}[t]
\begin{center}
\sidecaption
\includegraphics[scale=.4]{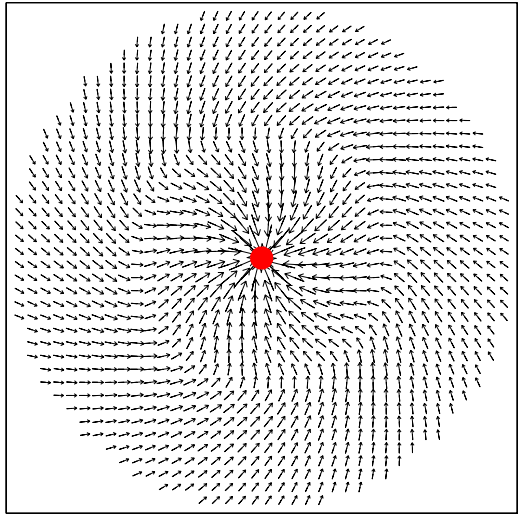}
\caption{
Illustration of the velocity field of the accretion flow in the
equatorial plane of a magnetized rotating cloud. In this figure the
inner solution of Galli et al. (2006) has been matched to the outer
rotating flow.  Because the azimuthal velocity decreases as $u_\varphi
\propto -j r^{1/2}$ for $r \rightarrow 0$, where $j$ is the gas
specific angular momentum at large distance, no rotationally supported
disk is formed. The flow shows a counter rotation at the center,
before the gas falls into the star. The magnetic field lines coincide
with  the streamlines outlined by the arrows.}
\label{vfield}   
\end{center}
\end{figure}

Galli et al.~(2006) studied the self-similar collapse of an
axisymmetric isothermal magnetized rotating cloud in the ideal MHD
regime and found an analytic solution that asymptotically approaches
free fall onto a central mass point, with an angular distribution
that depends on the mass loading of magnetic field lines. They found
that, independent on the details of the starting state, the magnetic
field acquires a split-monopole configuration where the magnetic
field is almost radial and directed in opposite directions above
and below the mid plane. In this configuration the radial magnetic
field strength increases as the inverse square of the distance 
$r$ from the origin,
\begin{equation}
|B_r |= \phi_*{c_s^3 t \over G^{1/2} r^2},
\label{Br}
\end{equation}
where $\phi_*$ is the non-dimensional magnetic flux trapped in the
central source, $c_s$ is the sound speed and $t$ is the time since
the onset of collapse.  This strong field produces a very efficient
magnetic braking that prevents the formation of a RSD.  The azimuthal
velocity of the infalling gas decreases to zero at the center as
$u_\varphi \propto - j r^{1/2}$, where $j$ is the specific angular
momentum of the gas in the envelope. Thus, the gas spirals into the
star with velocity approaching free-fall, $u_r \propto r^{-1/2}$.
The negative sign in the azimuthal velocity indicates that the
magnetic braking is so efficient as to enforce counter rotation in
the infalling gas, very close to the protostar. This counter rotation
of the innermost parts of the accretion flow has also been found
in numerical simulations (Mellon \& Li 2009; Kransnopolsky et al.~2010).  
The azimuthal component of the magnetic field decreases as
$B_\varphi \propto r^{-1}$ that increases with decreasing radius
more slowly than the poloidal component given by Equation \ref{Br}.
Thus, the winding of the field goes to zero near the protostar.
Figure \ref{vfield} illustrates the velocity field of the accretion
flow in the equatorial plane of a magnetized rotating cloud.  In
this figure the inner solution of Galli et al. (2006) has been
matched {\it ad hoc} to the outer rotating flow. The flow shows
counter rotation at the center, before the gas falls onto the star.

Summarizing, in the absence of magnetic torques, the angular momentum
of infalling fluid elements is conserved and a RSD is formed inside
a radius $r_d$, where the azimuthal velocity, increasing as $r^{-1}$,
becomes equal to the Keplerian velocity around the protostar,
increasing like $r^{-1/2}$. Instead, when magnetic braking dominates
over angular momentum conservation, the azimuthal velocity goes to
zero at small radii, and no RSD is formed.  Galli et al. named this
process ``catastrophic magnetic braking'' and concluded that the
dissipation of dynamically important levels of magnetic field is a
fundamental requisite for the formation of protoplanetary disks
around young stars.

Several numerical simulations addressed the question of determining
the maximum level of magnetization required to allow disk formation
under field-freezing conditions.  These studies have found that
disk formation is possible only for clouds with mass-to-flux ratios
$\lambda > 10$---$80$ (Mellon \& Li 2008; Hennebelle \& Fromang
2008; Seifried et al. 2011), or  $\lambda > 3$ for misaligned
magnetic and rotation axis (Hennebelle \& Ciardi 2009; Joos et al.
2012).  Nevertheless, misalignment alone is unlikely to solve the problem
of catastrophic magnetic braking at least for cores with $\lambda \approx 2$ 
(Li et al.~2013). Krumholz et al. (2013) recently proposed that a combination of
misalignment {\it and} weakness of the magnetic field may lead to the formation 
of disks in agreement with the observed fractions around embedded protostars. 
However, since the mass-to-flux 
ratio in molecular clouds is $\lambda \sim 1$--$4$ as discussed in \S \ref{sec:Intro}, it is clear the
constraint of ideal MHD must be relaxed at some point to allow the
formation of rotationally (rather than magnetically) supported disks.

Shu et al. (2006) solved analytically the problem of gravitational
collapse of a magnetized cloud with a uniform resistivity $\eta$,
in the kinematic approximation,  i.e. neglecting the back
reaction of the magnetic field on the motion of the gas. 
Under this simplifying assumption, dissipation of the magnetic field
occurs inside a sphere with radius $r_{\rm Ohm}$, that is inversely
proportional to the instantaneous stellar mass $M_*(t)$, and 
proportional to the square of the electrical resistivity $\eta$,
which may include Ohmic dissipation and AD:
\begin{equation}
r_{\rm Ohm} \equiv \frac{\eta^2}{2GM_*}
\approx 10\left(\frac{\eta}{10^{20}~\mbox{cm$^2$~s$^{-1}$}}\right)^2
\left(\frac{M_*}{M_\odot}\right)^{-1}~\mbox{AU}.
\end{equation}
Shu et al. showed that for values of $\eta \sim 10^{20}$~cm$^2$~s$^{-1}$,
the strength of the uniform magnetic field around the accreting
protostar is $B \sim 1$~G, of the order of the measured values in
meteorites. This anomalous resistivity is a few orders of magnitude
larger than the microscopic resistivity found by Nakano et al.
(2002) for the densities and ionization conditions of pseudo-disks
(see also Li et al. 2011). Kransnopolsky et al. (2010) performed
axisymmetric numerical simulations of resistive MHD relaxing the
kinematic approximation, and found that lower values of the resistivity
$\eta \sim 10^{19}$~cm$^2$~s$^{-1}$ were enough to allow the
formation of RSDs. Shu et al. also pointed out that the luminosity
resulting from the Ohmic dissipation of the electric current can
be very large,
\begin{equation}
L_{\rm Ohm} \sim 300\, \lambda^{-2} \left({M_*\over M_\odot}\right)^5 
\left({\eta \over 10^{20} \,{\rm cm^2 \, s^{-1}}}\right)^{-5} L_\odot.
\label{Ohmdiss}
\end{equation}
In fact, the resistivity cannot be much smaller than the values
quoted above, without violating the constraints on the observed
protostellar luminosity.

\begin{figure}[b]
\begin{center}
\sidecaption
\includegraphics[scale=.28]{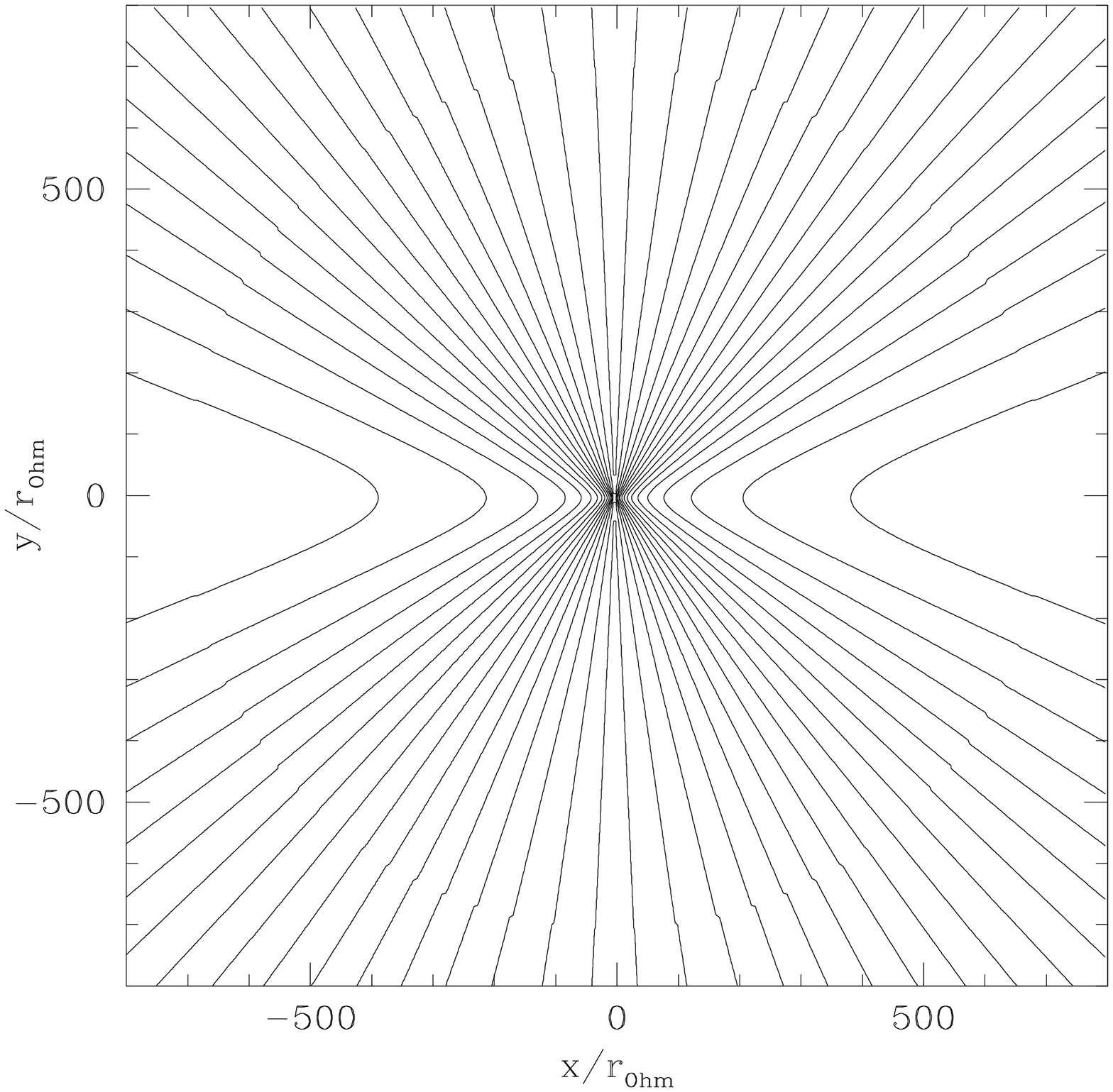}
\includegraphics[scale=.28]{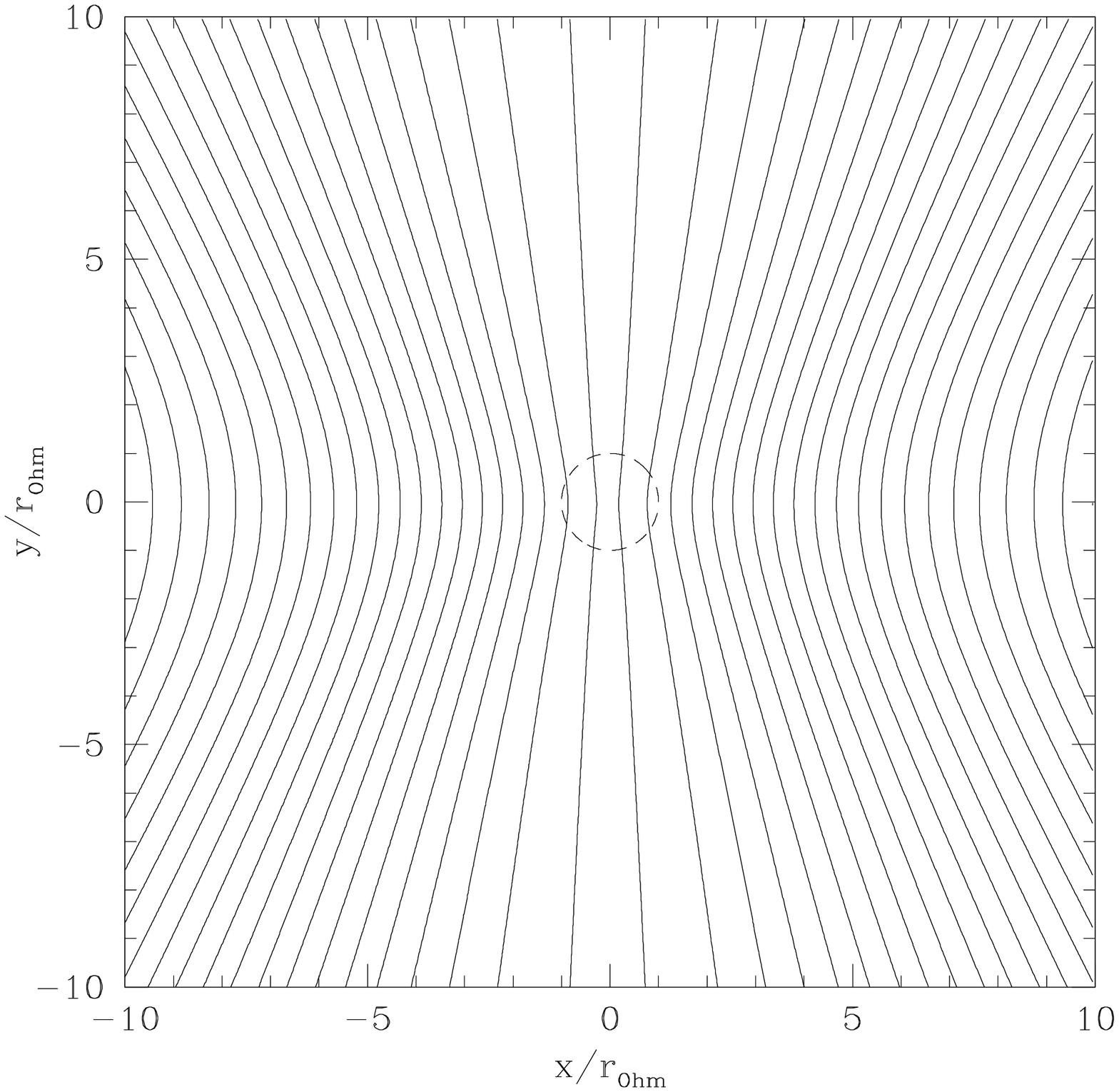}
\caption{Magnetic field configuration in the meridional plane for
magnetized cloud collapse with uniform resistivity. Distances are
measured in units of the Ohm radius $r_{\rm Ohm}$. The left panel
shows the field lines at large distance from the central source,
inside the region of radius $\le 500\, r_{\rm Ohm}$.  The right
panel shows the field lines inside a region of radius $\le10\,r_{\rm
Ohm}$ from the center.  The dashed circle has a radius equal to
$r_{\rm Ohm}$ (Figure from Galli et al.~2009).}
\label{ohm_radius}    
\end{center}
\end{figure}

Krasnopolsky \& K\"onigl (2002) and Braiding \& Wardle (2011) studied
self-similar 1-D models of the gravitational collapse of a flattened
rotating, magnetized cloud. Krasnopolsky \& K\"onigl found that
magnetic flux piles up at the so-called ``AD shock'' increasing the
efficiency of magnetic braking. Braiding \& Wardle included also
the Hall effect and found that RSDs could form for appropriate
values of the Hall coefficient and orientation of the magnetic
field, since the Hall diffusion is not invariant under field reversal.
However, in these self-similar models the diffusion coefficients
scale as $c_s^2 t$.  Thus, at typical times, $t \sim 10^{4-5}$~yr,
and for $c_s \sim 0.2$~km~s$^{-1}$, they have values $\sim
10^{20}$--$10^{21}$~cm$^2$~${\rm s}^{-1}$, larger than the expected
microscopic values, in agreement with Shu et al. (2006).

In recent years, several MHD simulations have been carried out
that include different non-ideal processes: Ohmic dissipation, AD,
and the Hall effect (see, e.g., Mellon \& Li 2009; Krasnopolsky et
al. 2010, 2011; Li et al. 2011).  These simulations find that, for
realistic levels of cloud magnetization and cosmic ionization rates,
flux redistribution by AD is not enough to allow the formation of
a RSD; and, as discussed above, in some cases, AD can even enhance
the magnetic braking.  They also require anomalous Ohmic resistivity
to form RSDs, and find that the Hall effect can spin up the gas
even in the case of an initially non-rotating cloud, although the
disks formed this way are sub-Keplerian.  These authors conclude that  the
combined effects of these diffusion mechanisms may weaken the
magnetic braking enough to form RSDs.

Machida et al.~(2011) performed 3-D resistive MHD simulations
of the collapse of strongly magnetized clouds ($\lambda \approx
$1--3).  Keeping the magnetic field anchored in the cloud's envelope, 
they found that the efficiency of magnetic braking 
depends on disk/envelope mass ratio:  magnetic braking is very efficient 
during the main accretion phase when the envelope is much more massive than  
the forming disk and the disk radius remains 
$\sim $10~AU or less;  magnetic braking then becomes largely ineffective 
when most of the envelope mass has been accreted by the star/disk system. 
As a consequence, the disk expands to a radius of $\sim 100$~AU during the 
late accretion phase, when the residual envelope mass is reduced to 
$\sim$ 70--80\% of the initial value (for a low-mass core).
This work suggests that Class 0 sources would have small
disks, while older sources would develop the observed hundred AU
disks, a prediction that could be tested by the new generation
interferometer ALMA.  In contrast with all other ideal MHD simulations,
Machida et al. also find the formation of a RSD in the ideal MHD
regime. A convergence study is necessary to asses the possible
influence of numerical diffusion in this result.  Also Dapp \& Basu
(2010) and Dapp et al.~(2012) find that Ohmic dissipation with
normal microscopic values is effective at high densities and RSDs
can form.  Nevertheless, their simulations address very early times,
when the disk has a size of only several stellar radii around a
protostar with mass $\sim 10^{-2}\, M_\odot$.  In particular, their
simulations are stopped when the expansion wave is only 15 AU away.
This is too early in the protostellar evolution process to determine
the possibility that the disk can survive through all the accretion
phase.

Another diffusion mechanism that has been proposed recently is the
reconnection diffusion of the magnetic field in a turbulent medium
(Lazarian \& Vishniac 1999).  In particular, Santos-Lima et al. (2012)
performed collapse simulations to study this process and find a
disk that is rotationally supported at radii between $\sim 65$--120
AU. Seifried et al. (2012) also included 
turbulence in ideal MHD simulations and claimed that this effect alone 
can lead to the formation of RSDs, without the requirement of magnetic flux loss. 
Nevertheless, these authors acknowledge a very high numerical diffusion in the region of disk
formation that, as discussed above, probably accounts for the field
dissipation and the weakening of the magnetic braking.
Furthermore,  Santos-Lima et al. (2013) argued that their apparent constant
mass-to-flux ratio is due to their averaging  over a large volume which can mask
the flux loss. On the other hand, it should also be kept in mind that observations 
indicate that dense cores are quiescent with the presence of only
subsonic turbulence (e.g., Pineda et al. 2010).  Therefore, any proposed 
mechanism of field diffusion should be able to work efficiently in conditions 
of weak turbulence. Also, since the small scales of the actual process of  
field reconnection cannot be resolved in the numerical simulations, the 
authors rely on numerical diffusion to mimic this physical process. We thus
consider that further tests and, possibly, semi-analytic studies are necessary 
to determine the efficiency of reconnection diffusion and the physical properties of 
disks produced by this mechanism.

Finally, it is interesting to notice that
several numerical simulations have found that the efficient  braking
and extra support provided by magnetic fields inhibit cloud
fragmentation  (e.g., Hosking \& Whitworth 2004; Hennebelle \&
Teyssier 2008; Duffin \& Pudritz 2009; Commer\c con et al. 2010,
2011). Also, the MHD numerical simulations that have been carried out to study
the problem of magnetic field diffusion assume in general an
isothermal equation of state and do not compute the energy released
by field dissipation. This power could be appreciable (see Equation
\ref{Ohmdiss}), and would be available to heat the gas and the dust,
accelerate particles, and thus, increase the ionization. It would be
important to include these effects in a self-consistent way in 
current models (e.g., Padovani et al. 2009; Glassgold et
al. 2012).  

In conclusion, how observed RSDs are formed around protostars
is still an open question.  The answer is probably a combination
of diffusive processes at high densities plus the dissipation of
the core envelope where the angular momentum is deposited by magnetic
braking.

\subsection{Measurements of Mass-to-Flux Ratios in Protostellar Disks}
\label{sec:ossdisks}

\begin{figure}[b]
\begin{center}
\sidecaption
\includegraphics[scale=.4]{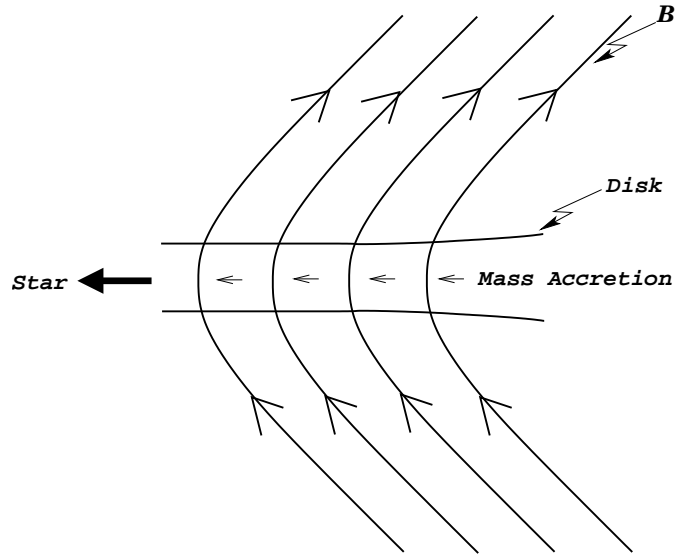}
\caption{Sketch of magnetized disk where the field lines are dragged by the accretion flow onto the star 
(from Shu et al. 2007).}
\label{Bdisk}    
\end{center}
\end{figure}

It is possible to estimate the values of the mass-to-flux ratio of
some star-disk systems and compare them to the measured values in
molecular cloud cores. This comparison provides an observational
constraint on the amount of magnetic flux dragged by the disk.  At
very large disk radii, magnetic field strengths of the order of a
few mG have been measured by Zeeman splitting in OH maser rings
associated to high-mass protostars (see Table~\ref{tab:lambda} for
references).  These ringlike configurations have radii of order
$10^3$~AU, are elongated in the direction perpendicular to the
outflow, and are usually characterized by a linear velocity gradient,
suggestive of rotation or accelerated expansion (see Cesaroni et al.~2007 for
a review).  In general, these observations suggest that, at these
radii, the field is mostly poloidal, although the presence of
reversals might indicate the presence of a toroidal component
probably generated by rotation.  Assuming that these measurements
actually probe a disk field and that it is possible to estimate the
system mass from the observed kinematics of the maser rings, one
can obtain the mass-to-flux ratio of the system.  If the vertical
component of the field $B_z$ scales with radius like $B_z(\varpi)\propto
\varpi^{-(1+\alpha)}$, with $\alpha\approx 3/8$ (Shu et al. 2007;
see also discussion in \S\ref{Bdisk}), most of the disk magnetic
flux $\Phi_{\rm d}$ is at large radii, $\Phi_{\rm
d}=2/(1-\alpha)B_z(\varpi)\varpi^2$.  The enclosed mass of disk
plus star can be estimated from the observed rotation, such that
$G(M_{\rm d}+M_*) \approx u_\varphi^2(\varpi)\varpi$. With these
assumptions, the mass-to-flux ratio of the system is
\begin{equation}
\lambda_{\rm sys} = \frac{2\pi G^{1/2} (M_*+M_{\rm d})}{\Phi_{\rm d}} \approx
(1-\alpha)\frac{u_\varphi^2(\varpi)}{G^{1/2} B_z(\varpi) \varpi}.
\end{equation}

Inserting the values of $B_z$, $u_\varphi$, and $\varpi$ measured
in OH maser rings around a few high-mass protostars (see
Table~\ref{tab:lambda}), we obtain values of $\lambda_{\rm sys}$
in the range 2--13.  Comparing these values of $\lambda_{\rm
sys}$ to the values $\lambda \approx 2$ typical of protostellar
cores, we conclude that the  mass-to-flux ratio of circumstellar disks 
around massive stars is somewhat larger, but not by a large factor, than 
the mass-to flux ratio of the parent cloud. In other words, the
magnetic flux trapped in circumstellar
disks around stars of mass $M_*\approx 10$~$M_\odot$ is only a
factor of a few smaller than the flux that would be trapped in the
system under field-freezing. These results, if confirmed, 
suggest that the solution to the catastrophic magnetic braking problem
discussed in Sect.~\ref{sec:magbrak} 
does not require a strong annihilation of magnetic field in the circumstellar
region (by, e.g., reconnection and/or turbulence), but rather a 
redistribution of the field (by a microscopic or macroscopic process) towards
a quasi-force-free configuration. In this respect, it will be very important to obtain the
mass-to-flux ratios also in disks around low-mass stars, an endeavor
that ALMA will make possible in the near future (see chapter by Vlemmings 
in this volume).

\begin{table}[t]
\caption{Estimates of the mass-to-flux ratio $\lambda_{\rm sys}$ in circumstellar disks 
around massive stars from OH Zeeman measurements.}
\label{tab:lambda}
\begin{tabular}{p{2cm}p{1.3cm}p{1.0cm}p{1.5cm}p{1.5cm}p{4.0cm}}
\hline\noalign{\smallskip}
source      & $u_{\varphi}$    & $B_z$ & $\varpi$ & $\lambda_{\rm sys}$ & reference \\
            & (km~s$^{-1})$    & (mG) & ($10^3$~AU) &           &      \\
\noalign{\smallskip}\svhline\noalign{\smallskip}
W57N        &  6  &  7  &  3   &  2.7  & Hutawarakorn et al.~(2002) \\
IRAS20126   &  3  & 11  & 0.85 &  1.5  & Edris et al.~(2007) \\
G35.2-0.74N &  5  &  5  & 2.6  &  3.1  & Hutawarakorn \& Cohen~(1999) \\
AFGL2591    &  5  &  4  & 0.75--1.5 & 6.7-13 & Hutawarakorn \& Cohen~(2005) \\
\noalign{\smallskip}\hline\noalign{\smallskip}
\end{tabular}
\end{table}

\section{Magnetized Accretion Disks }
\subsection{Viscosity and Resistivity}
\label{sec:Bdisk}

As discussed in the previous section, the RSDs drag a fraction of
the magnetic flux from the parent core during the gravitational
collapse.  Once the accretion has stopped, the magnetized disk will
evolve subject to two diffusive processes: viscosity, $\nu$, due
to turbulent and magnetic stresses, that produces accretion toward
the star and transfer of angular momentum outside; and resistivity,
$\eta$, due to microscopic collisions and the magnetorotational
instability (MRI; see, e.g., review of Balbus \& Hawley~1998), which
allows matter to slip across field lines. The MRI is considered
responsible for the disk ``anomalous'' viscosity needed to explain
the disks lifetimes of $\sim 10^6$~yr (Haisch et al. 2001;
Sicilia-Aguilar et al. 2004). As pointed out by Shu et al. (2007),
even in the case of a strong poloidal field, magnetized accretion
disks can develop the MRI instability.

In these disks, the dragging of field lines by accretion is balanced
by the outward field diffusion only if the ratio $\eta/\nu \sim
z_0/\varpi\ll 1$, where $z_0$ is the vertical half  disk thickness
and $\varpi$ is the radial cylindrical coordinate (Lubow et al.~1994).
This result is at variance with the usual assumption of a magnetic
Prandtl number $\nu/\eta \sim 1$ in  magnetized disks.  Moreover,
the magnetic tension due to the poloidal magnetic field threading
the disk will produce sub-Keplerian rotation (Shu et al. 2007).

In near field freezing conditions, the accretion flow generates a
mean radial field from the mean vertical field. This mean radial
field has two important consequences: First, it changes the radial
force balance and causes sub-Keplerian rotation of the gas.  If one
neglects the disk self-gravity and gas pressure, the force balance
equation is
\begin{equation}
\varpi \Omega^2 =  -{B_zB_\varpi^+\over 2\pi \Sigma} +{GM_*\over \varpi^2},
\label{centrifugal}
\end{equation}
where $\Omega$ is the rotation rate, $\Sigma$ is the disk mass
surface density, $B_z$ is the component of the magnetic field
threading vertically through the disk, and $B_\varpi^+$ is the
radial component of the magnetic field just above the disk that
responds to the radial accretion flow. The rotation rate, given by
the solution of the above equation, is smaller than the Keplerian
value $\left({GM_*/ \varpi^3}\right)^{1/2}$,
because of the extra support of the magnetic tension against
gravity,
\begin{equation}
\Omega = f \left({GM_*\over \varpi^3}\right)^{1/2}, \quad \mbox{with the sub-Keplerian factor $f < 1$}.
\end{equation}
Note that the field lines are bent because the sources of the disk
magnetization are currents at infinity anchoring magnetic field
lines to the parent cloud.  Second, the stretching of the poloidal
field by differential rotation produces an azimuthal field in the
disk that, coupled with the radial field, exerts a mean stress and
torques the gas, allowing the disk viscous evolution.  In the absence
of numerical simulations of the MRI that are both global and have
a non-zero net magnetic flux, Shu et al. (2007) proposed a
functional form of the viscosity based on mixing length arguments,
\begin{equation}
\nu=D\frac{B_z^2 z_0}{2\pi \Sigma \Omega},
\label{def_nu}
\end{equation}
where $D\le 1$ is a dimensionless  coefficient. Since, as discussed
above, the resistivity $\eta$ is related to viscosity in steady
state, Shu et al. were able to construct full radial models of disks
around young stars.  They find that the disk masses, sizes and
magnetic field strengths are consistent with the observations. The
coefficient $D$ should acquire small values if there are substantial
``dead zones'', where the ionization is too low to couple to magnetic
fields except, possibly, for thin surface layers (as could be the
case in disks around T Tauri stars). They propose that rapid transport
of mass and magnetic fluctuations across  strong mean field lines
can occur through the reconnection of small magnetic loops, twisted
and bent by the turbulent flow, this process being the source of
the disk viscous and resistive diffusivities. Since the microscopic
disk resistivities drop at large radii ($\varpi > 10$--20~AU; see,
e.g., Sano et al. 2000), Shu et al. conclude that the MRI has to
provide the anomalously large value of the resistivity in the outer
disk in order to allow accretion to continue to the central star.

For {\bf a} disk model with standard flaring ($z_0 \propto \varpi^{5/4}$)
Equation (\ref{centrifugal}) can be recast in the form
\begin{equation}
1-f^2 = {0.5444\over \lambda_{\rm sys}^2 } \left({M_* \over M_d} \right).
\label{ft}
\end{equation}  
For a closed star plus disk system in which infall has ceased, the
mass-to-flux ratio $\lambda_{\rm sys}$ remains constant. Then, since
disk accretion decreases the disk mass $M_d$  relative to stellar
mass $M_*$, the departure from Keplerian rotation, $(1-f^2)$, must
grow with time.  This happens because viscosity drains mass from
the disk onto the star, while resistivity can only cause the
redistribution of flux within the disk but cannot change the total
flux, making $f$ decrease with time.  Thus, the disk becomes more
sub-Keplerian and magnetized with time.

\subsection{Sub-Keplerian Disk Rotation}
\label{sec:rotcurve}

Accretion disks threaded by a poloidal magnetic field lines  as shown in Figure 1.3, 
bent with respect to the vertical by angles larger than 30$^\circ$,
are candidates to produce disk winds
(Blandford \& Payne 1982; for a recent review on disk winds, see, e.g. Pudritz et al.~2007).
In fact, in the models of magnetized disks of Shu et al. (2007) this criterion is
comfortably satisfied (see their Table 1).
Nevertheless, sub-Keplerian rotation of the gas in accretion disks poses a problem
to disk wind models.
In order to launch winds, sub-Keplerian disk either have to be warm to overcome the
potential barrier, or they need a dynamically fast diffusion across
the magnetic field lines (Shu et al. 2008).

In the case of thermal launching, given the fractional deviation
from Keplerian rotation $f<1$, the gas needs to climb the local
potential barrier in order to be ejected magnetocentrifugally along
field lines. Thus, the gas requires a thermal speed
\begin{equation}
c_s^2 \approx {1\over 4}(1- f^2){GM_*\over \varpi}.
\end{equation}
One can write this condition as a constraint for the gas temperature in terms 
of the local escape speed, $u_{esc} = \sqrt{2GM_*/\varpi}$,
 \begin{equation}
T \approx 1.4 \times 10^6 \left(1- f^2 \right) \left({u_{\rm esc} \over 200 \, {\rm km \,s}^{-1}}\right)  K.
\end{equation}
Thus, for a deviation from Keplerian rotation as small as  $f=0.95$,
typical of protostellar disks, the gas temperature need to be as
large as $T \sim 1.4 \times 10^5 \, K$ to eject a disk wind that
can reach typical protostellar speeds of $\sim 200 \,{\rm km\,
s^{-1}}$. Since the disks around young stars are cold ($T <1000 \,
K$; e.g., D'Alessio et al.~1999) thermal launching is not a viable
mechanism for magnetocentrifugal disk winds.

Resistive launching is in fact used by current disk wind models
that use a large resistivities that allow the gas to diffuse
vertically across magnetic field lines to the launching point where
$f=1$ (see, e.g., Figure 5 of Ferreira \& Pelletier~1995 showing
$f$ as a function of height).  Nevertheless, fast diffusion also
occurs radially, and, as a result, these models have accretion
speeds of the order of the sound speed, $u_\varpi \sim c_s$. Such
large speeds imply too short accretion timescales
\begin{equation}
\tau_{\rm acc} = {\varpi \over u_\varpi} \sim 2,400 \left( {\varpi \over 100 \, {\rm AU}}\right) 
\left({c_s \over 0.2 \,{\rm km \, s}^{-1}  }\right)^{-1} \, {\rm yr},
\end{equation}
i.e., these models have too short disk lifetimes (see \S \ref{sec:Bdisk}).
 
Therefore, magnetocentrifugally-driven, cold, disk wind models need
to face the challenge imposed by the sub-Keplerian gas rotation due
to the support provided by magnetic tension by a dynamically important
poloidad field: disk winds cannot be launched thermally and diffusive
launching makes the disks short lived.

\subsection{Stability and Planet Formation}
\label{sec:stabdisks}

Gravitational instabilities in accretion disks around young stars
can grow in the nonlinear regime and produce secondary bodies within
the disk, such as brown dwarfs and giant planets. On the other hand,
if the growing perturbations saturate, the gravitational torques
can lead to redistribution of angular momentum and disk accretion.
In non magnetic disks, both processes require the onset of gravitational
instability, which, for axisymmetric perturbations, is determined
by the value of the Toomre parameter $Q_T$,
\begin{equation}
Q_T \equiv \frac{c_s \kappa}{\pi G \Sigma} \, , 
\label{qtoomre} 
\end{equation}
where 
$\kappa=\varpi^{-1} [\partial(\varpi^2\Omega)^2/\partial\varpi]^{1/2}$ 
is the epicyclic frequency (Toomre 1964).

In the presence of magnetic fields, however, the condition of
gravitational instability is modified. The modification to the Toomre 
criterion for a magnetized disk has been discussed previously in 
the Galactic context by several authors (e.g., see Elmegreen~1994). However 
the magnetic field in the Galaxy is likely to be dynamo-generated and 
mostly toroidal (see chapter by Beck, this volume), whereas in a circumstellar disk the magnetic field is 
expected to be mostly poloidal and dragged from the parent cloud. From a linear stability
analysis, Lizano et al.~(2010) derived the modified
Toomre $Q_M$ parameter for a disk threaded by a poloidal magnetic field, 
which provides the boundary of stability
for axisymmetric ($m=0$) perturbations, given by
\begin{equation}
Q_M = \frac{\Theta^{1/2}a\kappa}{\pi \epsilon G \Sigma_0} \, ,
\label{qmagnetic} 
\end{equation}
where 
\begin{equation}
\Theta \equiv 1+\frac{B_{z}^2 z_0}{2\pi \Sigma c_s^2}\, \mbox{~~~and~~~}
\epsilon \equiv 1 - \frac{1}{\lambda^2} \, , 
\label{Theta_eps}
\end{equation}
and $\lambda=2\pi G^{1/2}\Sigma/B_z$ is the local value of the
mass-to-flux ratio in the disk.  For $Q_M<1$, perturbations with
wavenumber between $k_\pm=k_{\rm max}(1\pm \sqrt{1-Q_M^2})$ are
unstable, with $k_{\rm max}=(\epsilon/\Theta) k_J$ being the
wavenumber of maximum growth, and $k_J=\pi G \Sigma/c_s^2$ the Jeans
wavenumber. Since $\epsilon/\Theta <1$, the effect of the magnetic
field is to increase the length scale of the gravitational instability
with respect to the Jeans length scale.

\begin{figure}[t]
\begin{center}
\sidecaption
\includegraphics[scale=.5]{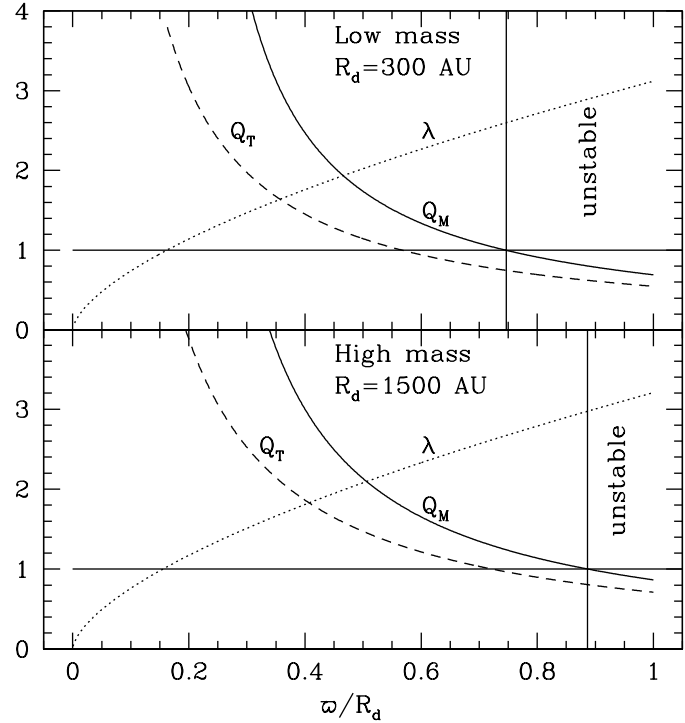}
\caption{Radial profiles of the Toomre stability parameter $Q_T$,
the magnetic stability parameter $Q_M$, and the local mass-to-flux
ratio $\lambda$ in typical disks around high-mass stars (lower
panel) and low-mass stars (upper panel). These profiles are obtained
from the disk models of Shu et al. (2007)  with an aspect ratio
$\propto \varpi^{1/4}$. The radial coordinate is normalized to the
disk radius $R_d$.}
\label{fig:low-high}    
\end{center}
\end{figure}

Another important factor that determines $Q_M$ in Equation
(\ref{qmagnetic}) is the epicyclic frequency. As discussed in
\S\ref{sec:Bdisk}, magnetized disks around young stars rotate at
sub-Keplerian speeds because magnetic tension modifies the force
balance equation, then $\kappa=f\Omega_K$.  Therefore, the inclusion
of magnetic fields produces competing effects on the instability
parameter $Q_M$: the strong fields enforce sub-Keplerian flow, which
reduces $Q_M$ and leads to greater instability; on the other hand,
both magnetic pressure and magnetic tension act to increase $Q_M$
and lead to enhanced stability. For typical disks around low- and
high-mass stars, the stabilizing effect wins in the inner regions,
as shown in Figure~\ref{fig:low-high}.  Thus, stable magnetized
disks can be more massive than their non-magnetized counterparts.
The two panels in Figure \ref{fig:low-high} show the values of the
parameters $Q_T$, $Q_M$ and the local mass-to-flux ratio $\lambda$,
for the disk models of Shu et al.~(2007) as function of the normalized
disk radius $\varpi/R_d$.  Since $Q_M$ is always larger than $Q_T$,
the magnetic field has a stabilizing effect against gravity, the
radius of the stable region increasing by $\sim 20$--$30$\% with
respect to a nonmagnetic disk.  Correspondingly, the fraction of
stable enclosed disk mass, $m(\varpi)/M_d$, where $M_d$ is the total
disk mass, increases by $\sim 30$--$40$\%.  Also, the unstable
region is magnetically supercritical ($\lambda >1$), as required
to allow local gravitational collapse.

The increased stability of magnetized disks against gravitational
perturbations poses an obstacle to the formation of giant planets,
or somewhat larger secondary bodies such as brown dwarfs, by this
process. As discussed by Lizano et al.~(2010), the formation of
giant planets in the outer parts of disks where $Q_M<q_*\approx 2$
has to satisfy, in addition, the condition of short cooling time
$\tau_{\rm cool}$, and the rapid loss of an appreciable amount of
magnetic flux. The coupled constraints on $Q_M$ and $\tau_{\rm
cool}$ result in a condition for the Minimum Mass Solar Nebula (MMSN) disk,
on the minimum radius
outside which giant planet formation can occur,
\begin{equation}
\varpi > 
1100 \left( f {\cal F} \right)^{2} \,  {\rm AU} ,  
\label{radialscale} 
\end{equation}
where ${\cal F}\approx 0.5$ is a nondimensional quantity that depends on the 
mass-to-flux ratio $\lambda$. The condition on the 
magnetic flux loss gives a constraint on the disk resistivity
\begin{equation}
\eta >  2.5 \times 10^{18} \,{\rm cm^2 \,s^{-1} },
\label{etacon}
\end{equation}
larger than the microscopic resistivity.

Therefore, magnetic fields stabilize protoplanetary disks, and make
it more difficult to form giant planets via gravitational instability.
At any rate, this process can occur only at large radii.

\subsection{Implications of Disk Magnetization for Planet Migration}
\label{sec:migraplan}

Protoplanets experience orbital migration due to angular momentum exchange
by tidal interactions with the disk material. There are several types of 
tidal interactions that occur through wave excitation or advected disk 
material: Type I migration applies to embedded protoplanets; Type II
migration occurs when the protoplanet is massive enough to open a gap;
and Type III migration is driven by coorbital torques (see review Papaloizou \& Terquem 2006).
In magnetized disks,  sub-Keplerian rotation results in a new migration mechanism
for embedded proto-planets  (Adams et al. 2009). These bodies
rotating at Keplerian speed, $\Omega_K = (G M_*/r^3)^{1/2}$, where
$r$ is the planet semi-major axis, experience a headwind against
the magnetically controlled gas that rotates at sub-Keplerian speeds.
The drag force drives their inward migration. The relative speed
between the gas and the proto-planet is $u_{\rm rel} = (1-f)\Omega_K
r $, and the torque is
\begin{equation}
T= C_D \pi R_P^2 r \rho_g u_{\rm rel}^2 =
{\pi \over 2} C_D \left(1-f \right)^2 \Omega_K^2 r^3 R_P^2 \left({\Sigma \over z_0} \right),
\end{equation}
where $C_D \sim 1$ is the drag coefficient, $R_P$ is the planet radius, 
$\rho_g = \Sigma/2 z_0$ is the gas density, and $z_0$ is the disk scale height. 
Assuming circular orbits,  the time change of the planet's angular momentum gives 
the time evolution of  $r$ 
\begin{equation}
{1\over r } {d r \over d t} = {2 T \over m_P \Omega_K r^2},
\end{equation}
that implies a migration timescale $t_M \sim 70,000$ yr for an
Earth-like planet in a MMSN disk with
$f\sim 0.66$.

\begin{figure}[t]
\begin{center}
\sidecaption
\includegraphics[scale=.4]{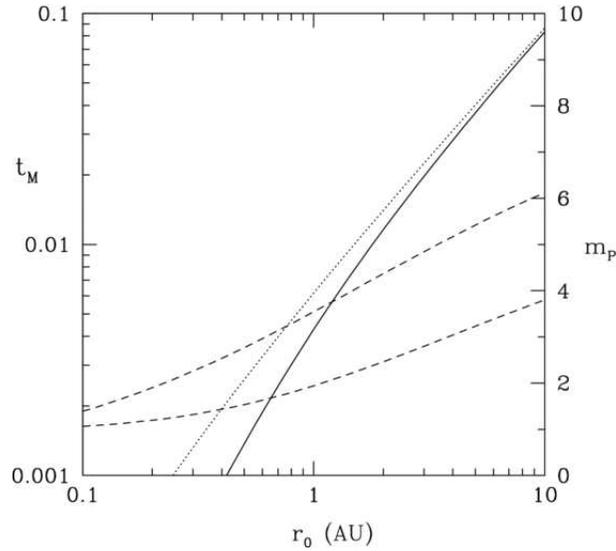}
\caption{Migration time and final core mass versus starting radius, $r_0$. 
The solid curve shows migration time $t_M$ in  Myr (left axis)
including sub-Keplerian and Type I torques. The dotted curve shows
$t_M$ for Type I torques only.  The dashed curves show final core
mass $m_P$ in $M_x$ (right axis) for migration with both torques
(bottom) and Type I torques only (top) (from Adams et al. 2009).}
\label{mig}       
\end{center}
\end{figure}

This mechanism dominates over Type I migration 
for sufficiently small planets with masses $m_P \le 1 M_{\rm Earth}$, and/or close orbits
$r < 1$~ AU.  Taking into account both mechanisms, the total migration
time $t_M$ moderately decreases due to the sub-Keplerian torques,
but the mass accreted by planetary cores during the migration epoch
changes more substantially, as shown in Figure \ref{mig}.
Furthermore, Paardekooper (2009) showed that disk-planet interactions
in Type I migration are affected by the degree of  disk
sub-Keplerian rotation because the position of the  Linblad and
corotation resonances change with the sub-Keplerian factor
$f$. Paardekooper concluded that
migration in sub-Keplerian disks, in general, will be directed
inwards and will be sped up compared to Keplerian disks.
In general, Type I migration is very fast, with timescales
of the order of $10^5$ yr, implying  that low-mass planets have a hard time 
surviving in gaseous disks; thus,  they would
need to form at later times, when the gas has been accreted or dispersed. 
Moreover, as discussed by Papaloizou \& Terquem (2006), 
even the survival of the rocky cores of giant planets would be compromised 
unless this type of migration is somehow suppressed or modified by, for example,
eccentricity effects (Papaloizou 2002), large scale toroidal magnetic fields
(Terquem 2003), stocastic  torques due to the presence of turbulence 
(Nelson and Papaloizou 2004; Laughlin et al. 2004; 
Adams \& Bloch 2009); or by disk opacity effects (Menou \& Goodman 2004; 
Paardekooper \& Mellema 2006).

\section{Conclusions}

Magnetic fields are dynamically important for star formation. The
measured levels of cloud magnetization, $\lambda=1$--4, imply that
the clouds are close to the critical value of the mass-to-flux ratio
to provide support against gravitational collapse. Also, observations
of polarized dust emission and statistical analysis of polarization
maps strongly suggest that the field is well ordered on pc scales,
and has only a relatively small turbulent component.

Although it has been believed since the 1970s that magnetic braking 
is responsible for the loss of angular momentum in cloud cores, 
an unexpected result has been
the theoretical finding that in an ideal MHD flow the magnetic
braking becomes so efficient as to prevent the formation of
centrifugally supported disks.  Therefore, magnetic field  diffusion and/or
dissipation
in the high density regime of gravitational collapse is needed to
avoid catastrophic braking. Field dissipation was also advocated
to solve the magnetic flux problem in newly born stars.  Several
non ideal MHD diffusion processes, able to 
redistribute the magnetic field brought by gravity in the collapse region,
have been studied in analytic and numerical
simulations to alleviate the catastrophic magnetic braking, but all
of them have been found relatively inefficient. Thus, at the moment,
the problem has not yet been resolved. Conversely, solutions 
based on magnetic reconnection may require high levels of turbulence,
which are generally not observed in dense cloud cores.

Once the centrifugally supported disks form, they are expected to
have important levels of magnetization due to the incomplete
diffusion and /or dissipation of the magnetic field dragged during the gravitational
collapse. The strong poloidal field produces sub-Keplerian rotation
of the gas because magnetic tension provides support against gravity.
This slower rotation poses a local potential barrier that disk winds
have to overcome, either by thermal pressure or by fast diffusivity,
to reach the launching point where magnetocentrifugal acceleration
can take place.  Since thermal launching is not possible in cold
disks around low-mass stars, disk wind models rely on fast diffusion
of the gas across field lines. However, this fast diffusion also
produces accretion flows with velocities of the order of the sound
speed, that will empty the disk into the star in very short timescales,
of only several thousand years. Up to now, no solution has been
proposed to this conundrum.  Sub-Keplerian rotation of the gas in
the disk also produces a headwind on forming proto-planets that
move at Keplerian speeds, making them lose angular momentum and
migrate faster toward the central star.

Finally, the magnetic fields modify the disk stability by two
opposing effects: magnetic pressure and tension support the gas
against gravitational collapse, but sub-Keplerian rotation makes
the gas locally more unstable.  The resulting magnetically modified
Toomre stability parameter, $Q_M$, is in general larger than its
nonmagnetic counterpart in accretion disks around young stars. Thus,
stable magnetized disks can be more massive that nonmagnetic disks.
The region of instability is pushed at larger radii, making it more
difficult to form giant planets via gravitational instability.

In the near future, ALMA will be able to measure magnetic fields
and disk rotation curves with unprecedented spatial resolution and
test  the theoretical models discussed here.  BLAST-pol will map
the magnetic field direction of a large sample of molecular clouds
and determine the relative strength of the large scale versus the
turbulent magnetic field components testing the importance of
magnetic fields in cloud support and evolution.

\begin{acknowledgement}
S. L. and D. G. acknowledge support from the Scientific Cooperation
Agreement Mexico-Italy MX11M07: ``The formation of disks and planets around young stars''. 
S. L. also acknowledges support from
PAPIIT-UNAM IN100412. The authors  thank Fred C. Adams, Anthony Allen,
Michael J. Cai, Alfred E. Glassgold, and Frank H. Shu for a longtime
enjoyable collaboration, and an anonymous referee for a detailed 
and thoughtful report.
\end{acknowledgement}


\end{document}